\documentclass[showkeys,showpacs,amssymb,amsmath,preprint]{revtex4}
\usepackage{amsmath}

\usepackage{amssymb}
\usepackage{graphicx}% Include figure files
\usepackage{dcolumn}% Align table columns on decimal point
\usepackage{bm}% bold math
\usepackage{setspace}
\usepackage{bbm}

\usepackage{amsfonts}

\usepackage{dcolumn}% Align table columns on decimal point
\usepackage{bm}% bold math
\usepackage{romannum}

\begin{document}
\title{Solving the generalized Higgs model from the generalized CRS model}
\author{Ci Song}
\email[Email:]{ cisong@mail.nankai.edu.cn} \affiliation{Theoretical
Physics Division, Chern Institute of Mathematics, Nankai University,
Tianjin, 300071, P.R.China
\\PHONE: 011+8622-2350-9287,
FAX: 011+8622-2350-1532}

\author{Yan Li}
\affiliation{Theoretical
Physics Division, Chern Institute of Mathematics, Nankai University,
Tianjin, 300071, P.R.China
\\PHONE: 011+8622-2350-9287,
FAX: 011+8622-2350-1532}

\author{Jing-Ling Chen}
\email[Email:]{ chenjl@nankai.edu.cn} \affiliation{Theoretical
Physics Division, Chern Institute of Mathematics, Nankai University,
Tianjin, 300071, P.R.China
\\PHONE: 011+8622-2350-9287,
FAX: 011+8622-2350-1532}

\date{\today}

\begin{abstract}
 In this paper, we reveal a direct relation between the generalized one-dimensional Carinena-Ranada-Santander (CRS) model and the radial part of two-dimensional generalized Higgs model. By this relation, we construct a series of quasi-exactly solutions for the two-dimensional Higgs model from a solved generalized CRS model.
\end{abstract}


%Fokas-Lagerstrom potential, Smorodinsky-Winternitz potential, Holt
%potential}

\maketitle

\section{Introduction}

In this paper, we present a coordinate transformation between the one-dimensional radial generalized Higgs model\cite{Higgs} and the generalized Carinena-Ranada-Santander (CRS) model \cite{CRS,wang}. Using this transformation, we construct a series of two-dimensional quasi-exactly solvable (QES) Higgs models.

The coordinate transformation is a useful method for construct physical models. Historically, Euler obtained the transformation between the equation of one-dimensional Kepler motion and the equation of a one-dimensional harmonic oscillator \cite{Euler}. Levi-Civit$\grave{a}$ reduced the two-dimensional Kepler problem to a two-dimensional harmonic oscillator \cite{LC}. Kustaanheimo and Stiefel got the reduction of the three-dimensional Kepler problem to a four-dimensional harmonic oscillator \cite{KS}. Davtyant et al showed how to transform from the five-dimensional Kepler problem to a eight-dimensional harmonic oscillator \cite{LS-LG}. Moreover, the coordinate transformation method can also be applied to many other problems, such as celestial mechanics \cite{SS}, the complicated classical dynamics of the hydrogen atom in crossed electric and magnetic fields \cite{GUF,von,SZ}, and similar problems in quantum mechanics \cite{FLZhang,Fubo,DBY}. These results have also been discussed in mathematical field \cite{Bartsch}.

The QES quantum problem was a remarkable discovery in last century \cite{QEM-1,QEM-2,QEM-3,QEM-5,QEM-6}. Quantum-mechanical potentials are quasi-exactly solvable if only parts of energy spectrum and associated eigenfunctions can be solved as exactly analytical form. QES systems are intermediate to non-exactly solvable and solvable ones and their potentials depend on a parameter. This kind of problem can be solved by Lie algebra \cite{LA-1}, analytical method \cite{AM-1}. Three methods for construction of QES potentials were used to establish many QES potentials, which are respectively based on an ansatz method, canonical transformation, and supersymmetric (SUSY) quantum mechanics.

Higgs introduced a generalization of the isotropic harmonic oscillator and hydrogen atom on a two-dimensional curved sphere with constant curvature $\lambda_G$ \cite{Higgs}. Leemon generalized it into N-dimensional curved sphere \cite{Leemon}. On a two-dimensional curved sphere, the Hamiltonian can be written as
\begin{equation}\label{eq:Higgs}
H_G=\frac{1}{2m}\left(\pi^2+\lambda_GL^2\right)+\mathcal{V}(r),
\end{equation}
where $\pi=\mathbf{p}+\frac{1}{2}\lambda_G\left[\mathbf{x}\left(\mathbf{x}\cdot\mathbf{p}\right)+\left(\mathbf{p}\cdot\mathbf{x}\right)\mathbf{x}\right]$, $L^2=\frac{1}{2}L_{ij}L_{ij}$, $r=\left|\sqrt{\mathbf{x}^2}\right|$, $\mathcal{V}(r)=\frac{1}{2}m\omega^2r^2$ for the harmonic oscillator, and $\mathcal{V}(r)=-\frac{\kappa}{r}$ for the hydrogen atom. Carinena, Ranada and Santander proposed a kind of one-dimensional model for the quantum nonlinear harmonic oscillator, which is called the CRS model now \cite{CRS}. Wang and Liu generalized a class of the exactly solvable CRS model \cite{wang}, whose Hamiltonian reads
\begin{equation}\label{eq:wang}
H_Q=\frac{1}{2m}\left(\mathcal{K}p^2-i\hbar\lambda_Qxp\right)+V(x),
\end{equation}
where $\mathcal{K}=1+\lambda_Qx^2$, $\lambda_Q$ is a real number, and the potential $V(x)$ is determined by the factorization method as $V(x)=\frac{\hbar^2}{2m}\cdot\frac{\left(\beta X+\gamma\right)^2+\left(\beta X+\gamma\right)\left(AX+B\right)}{\mathcal{K}\left(\frac{dX}{dx}\right)^2}+C$.
For the potential $V(x)$, $\beta$, $\gamma$ and $C$ are arbitrary numbers. $X=X(x)$ is a function which is analytic nearby $x=0$. The parameters $A$ and $B$ need to satisfy the equation $\mathcal{K}\frac{d^2X}{dx^2}+\lambda_Qx\frac{dX}{dx}=AX+B$.

In this paper, we find a coordinate transformation between the radial generalized Higgs model and the generalized CRS model. According to this transformation, each exactly solvable generalized CRS model corresponds to an exactly solvable radial generalized Higgs model. However, for each of these exact solvable radial generalized Higgs model, the fixed parameter $m'_Q$ appearing in the transformed potential $\mathcal{V}(r)$ needs to equal to the angular parameter $m'$ of the two-dimensional generalized Higgs model. Thus, only a part of exact solvable radial generalized Higgs model can be solved on the two dimensions.

The paper is organized as follows. In Sec. 2, this transformation is given; in Sec. 3, the transformed Higgs models which are quasi-exactly solvable will be shown; in Sec. 4, there is a conclusion finally.

\section{The transformation between generalized CRS model and the radial generalized Higgs model}

In this section, we give the coordinate transformation between the radial part of the generalized Higgs model and the generalized CRS model.

\subsection{The exact solution of the Higgs oscillator}

First, we calculate the wave-function and the energy spectrums of the Higgs oscillator.

The stationary Schr\"{o}dinger equtaion  of the two-dimensional Higgs oscillator with the potential $\mathcal{V}(r)$  $\frac{1}{2}m\omega^2r^2$ is
\begin{eqnarray}\label{eq:Higgs-par-eq}
&&-\frac{\hbar^2}{2m}\left[(1+\lambda_Gr^2)^2\frac{\partial^2}{\partial r^2}+\frac{(1+\lambda_Gr^2)(1+5\lambda_Gr^2)}{r}\frac{\partial}{\partial r}+\left(3\lambda_G+\frac{15\lambda_G^2r^2}{4}\right)+(\lambda_G+\frac{1}{r^2})\frac{\partial^2}{\partial\theta^2}\right]\Psi(r,\theta)\nonumber\\
&=&(E_G-\frac{1}{2}m\omega^2r^2)\Psi(r,\theta),
\end{eqnarray}
where $E_G$ is the stationary energy eigenvalue. Considering $\Psi(r,\theta)=e^{im'\theta}\psi(r)$, we can seperate the angular part from the equation above and get the radial part of above equation
\begin{eqnarray}\label{eq:Higgs-rad-eq}
&&-\frac{\hbar^2}{2m}\left[(1+\lambda_Gr^2)^2\frac{d^2}{dr^2}+\frac{(1+\lambda_Gr^2)(1+5\lambda_Gr^2)}{r}\frac{d}{dr}+\left(3\lambda_G-\lambda_Gm'^2+\frac{15}{4}\lambda_G^2r^2-\frac{m'^2}{r^2}\right)\right]\psi(r)\nonumber\\
&=&(E_G-\frac{1}{2}m\omega^2r^2)\psi(r).
\end{eqnarray}
By solving this radial equation, we get the radial wave function
\begin{equation}\label{eq:Higgs-rad-wf}
\psi(r)_{N,m'}=r^{|m'|}\left(\frac{1}{1+\lambda_Gr^2}\right)^{1+\frac{|m'|}{2}+\frac{m\omega_G'}{2\hbar\lambda_G}}{{_2}F_1}(-N,N+|m'|+1+\frac{m\omega_G'}{\lambda_G\hbar},|m'|+1;\frac{\lambda_Gr^2}{1+\lambda_Gr^2})
\end{equation}
and the energy spectrum
\begin{equation}\label{eq:Higgs-energy}
E_{G(N,m')}=\hbar\omega_G'(2N+|m'|+1)+\frac{\lambda_G\hbar^2}{2m}(2N+|m'|+1)^2,
\end{equation}
where $\omega'_G=\sqrt{\omega^2+\frac{\hbar^2\lambda_G^2}{4m^2}}$, $N$ and $m'$ are both integer numbers.

\subsection{The exactly solution of a special generalized CRS model}

In this part, we calculate the wave function and the energy spectrums of a special generalized CRS model.

For the generalized CRS model, if we set the function and parameters in the potential as $X(x)=\cos\left(2\Theta(x)\right)$, $\beta=2\lambda_Q(m'_Q+1)+\sqrt{\lambda_Q^2+\frac{4m^2\omega^2}{\hbar^2}}$, $\gamma=2\lambda_Qm'_Q-\sqrt{\lambda_Q^2+\frac{4m^2\omega^2}{\hbar^2}}$, $A=-4\lambda_Q$, $B=0$, and $C=\frac{\hbar^2}{2m}\left(\lambda_Q({m'_Q}^2-1)+m'_Q\sqrt{\lambda_Q^2+\frac{4m^2\omega^2}{\hbar^2}}\right)$, we get
\begin{equation}\label{eq:potential-exact}
V(x)=\frac{1}{2}m\omega^2\left(\frac{\tan(\Theta(x))}{\sqrt{\lambda_Q}}\right)^2-\frac{\lambda_Q\hbar^2}{8m}\left(1+(1-4{m'_Q}^2)\csc^2(\Theta(x))\right),
\end{equation}
where $\Theta(x)=\textrm{arcsinh}\left(\sqrt{\lambda_Q}x\right)$. With the potential above, by solving the generalized CRS eigen-equation
\begin{equation}\label{eq:CRS-eigen-eq}
\left[\frac{\hbar^2}{2m}\left(-\mathcal{K}\frac{d^2}{dx^2}-\lambda_Qx\frac{d}{dx}\right)+V(x)\right]\phi(x)=E_Q\phi(x),
\end{equation}
we get the wave function
\begin{eqnarray}\label{eq:CRS-wf}
\phi(x)&=&\left[-\sin^2(2\Theta(x))\right]^{-\frac{3}{4}}\sin^2(\Theta(x))\left(\frac{\tan(\Theta(x))}{\sqrt{\lambda_Q}}\right)^{|m'_Q|}\\
&&\left[\cos(\Theta(x))\right]^{1+\frac{|m'_Q|}{2}+\frac{m\omega_Q'}{2\hbar\lambda_Q}}{{_2}F_1}(-N,N+|m'_Q|+1+\frac{m\omega'_Q}{\lambda_Q\hbar},|m'_Q|+1;\sin(\Theta(x)))\nonumber
\end{eqnarray}
and the energy spectrum
\begin{equation}\label{eq:CRS-energy}
E_{Q(N,m'_Q)}=\hbar\omega_Q'(2N+|m'_Q|+1)+\frac{\lambda_Q\hbar^2}{2m}(2N+|m'_Q|+1)^2,
\end{equation}
where $\omega_Q'=\sqrt{\omega^2+\frac{\hbar^2\lambda_Q^2}{4m^2}}$, $N$ is integer number and $m'_Q=\frac{\beta+\gamma}{4\lambda_Q}-\frac{1}{2}$.

\subsection{The transformation relation between the generalized CRS model and the radial generalized Higgs model}

In this section, we give the transformation relation between the generalized CRS model and the radial generalized Higgs model.

Firstly, there exists two conditions of this coordinate transformation, which are respectively $\lambda_Q=\lambda_G=\lambda$ and the angular parameter $m'$ equaling to the fixed parameter $m'_Q$ which appears in the generalized CRS model. With both of these conditions, comparing the energy spectrums (\ref{eq:Higgs-energy}) and (\ref{eq:CRS-energy}), it is obviously that they are exactly the same. Thus, we achieve the coordinate transformation
\begin{equation}\label{eq:transformation}
x(r)=\frac{1}{\sqrt{\lambda}}\sinh(\Upsilon(r)),\ \ \ \ \Upsilon(r)=\arctan(\sqrt{\lambda}r),
\end{equation}
meanwhile, we get the same form of wave-function (\ref{eq:Higgs-rad-wf}) as (\ref{eq:CRS-wf}) from the following relation
\begin{equation}\label{eq:transformation-wf}
\psi(r)=g(r)\phi(x(r)),\ \ \ \ \ \ \ g(r)=-2(1-i)\left(\lambda r^2\right)^{-1/4}(1+\lambda r^2)^{-1/2}.
\end{equation}
Therefore, we get the same differential equation (\ref{eq:Higgs-rad-eq}) as (\ref{eq:CRS-eigen-eq}), in which $V(x(r))$ and $\mathcal{V}(r)$ satisfies the following relation of the potentials
\begin{equation}\label{eq:transformation-V}
\mathcal{V}(r)=V(x(r))+\frac{\lambda\hbar^2}{8m}\left[1+(1-4{m'_Q}^2)\left(1+\frac{1}{\lambda r^2}\right)\right].
\end{equation}

Here, we need the parameters of the generalized CRS model satisfy $\beta=2\lambda(m_Q'+1)+\lambda\delta$, $\gamma=2\lambda m_Q'-\lambda\delta$ and $C=\frac{\hbar^2}{2m}\left(\lambda({m_Q'}^2-1)+m_Q'\lambda\delta\right)$, where $\delta=\sqrt{1+\frac{4m^2\omega^2}{\lambda^2\hbar^2}}$ .

\section{The QES generalized Higgs model}

In this section, we will reveal how to get the QES generalized Higgs model and give some explicit examples.

By observing the transformation between the generalized CRS model and the radial Higgs model, it is easy to find that the transformed Higgs potential $\mathcal{V}(r)$ (\ref{eq:transformation-V}) in the two-dimensional Higgs model cannot actually be solved exactly for each of the corresponding exactly solvable generalized CRS model. In this transformed potential  (\ref{eq:transformation-V}), the parameter $m'_Q$ is fixed by the parameter $\lambda$, $\beta$ and $\gamma$. Meanwhile, the transformation requires the angular parameter $m'$ of the generalized Higgs model to be equal to the parameter $m'_Q$. When $X(x)=\cos\left(2\Theta(x)\right)$, this transformation has just canceled the parameter $m'_Q$ in the transformed potential $\mathcal{V}(r)$. For a general case, however, the parameter $m'_Q$ in potential $\mathcal{V}(r)$ cannot be directly separated from the radial coordinate $r$ for angular parameter $m'$. So, it is obvious that this transformed Higgs model have the exact solution  $\Psi(r,\theta)=e^{im'_Q\theta}\psi(r)$ in the case of $m'=m'_Q$, and the other solution for angular parameter $m'\neq m'_Q$ can not be gotten analytically. Thus, we can get a series of the QES Higgs models from the exactly solvable generalized CRS models.

In Ref. \cite{wang}, for each exactly solvable CRS model, the analytical function $X(x)$ needs to satisfy the equation $\mathcal{K}\frac{d^2X}{dx^2}+\lambda_Qx\frac{dX}{dx}=AX+B$. Thus, there exists a general solution
\begin{equation}\label{eq:X(x)}
X(x)=-\frac{B}{A}+C_1\cosh\left(\sqrt{\frac{A}{\lambda}}\text{arcsinh}\left(\sqrt{\lambda}x\right)\right)+iC_2\sinh\left(\sqrt{\frac{A}{\lambda}}\text{arcsinh}\left(\sqrt{\lambda}x\right)\right),
\end{equation}
which gives a solvable generalized CRS model for arbitrary $A$, $B$, $C_1$ and $C_2$ and give a corresponding QES generalized Higgs model simultaneously.\ \\

\textbf{Example 1. $A=-\lambda l^2$, $B=0$, $C_1=1$, $C_2=0$}

In this case, $X(x)=\cos\left(l\cdot\textrm{arcsinh}\left(\sqrt{\lambda}x\right)\right)$. By the above transformation, we get the QES potential
\begin{eqnarray}\label{eq:V-X=l}
\mathcal{V}(r)&=&\frac{\hbar^2}{8mr^2}\left[1-4{m_Q'}^2+2\lambda r^2(4{m_Q'}+3)+4\lambda r^2({m_Q'}+1)\delta\right]\\
&&-\frac{\lambda\hbar^2}{4ml^2}\left[10+8{m_Q'}({m_Q'}+2)+8({m_Q'}+1)\delta+(l^2-4{m_Q'}-2)(2{m_Q'}+1)\csc^2(\frac{l}{2}\Upsilon(r))\right.\nonumber\\
&&\ \ \ \ \ \ \ \ \ \ \ \left.+(l^2-4)\left(1+\delta\right)\sec^2(\frac{l}{2}\Upsilon(r))\right]\nonumber\\
&&+\frac{2}{l^2}m\omega^2\left[\frac{\tan\left(\frac{l}{2}\Upsilon(r)\right)}{\sqrt{\lambda}}\right]^2\nonumber
\end{eqnarray}
and the ground-state wave function
\begin{equation}\label{eq:f-X=l}
\psi_0(r)=\mathcal{N}_0(\lambda r^2)^{-1/4}(1+\lambda r^2)^{-1/2}\left[\tan\left(\frac{l}{2}\Upsilon(r)\right)\right]^{\gamma/(\lambda l^2)}\left[\sin\left(\frac{l}{2}\Upsilon(r)\right)\right]^{\beta/(\lambda l^2)}.
\end{equation}

The excited state wave functions can be gotten by the raising operator $b^\dagger$ in Ref. \cite{wang} $\phi_n(x)=(b^\dagger)^n\phi_0(x)$ and the transformation as $\psi_n(r)=g(r)\phi_n(x(r))$.

For the excited state, however, the two-dimensional generalized Higgs model has exactly solutions only in the condition of angular parameter $m'$ equaling to $m_Q'$. In other angular parameter cases, it cannot be  solved. Therefore, it is a quasi-exactly solvable model.

For $l=2$, we get the harmonic oscillator from the potential (\ref{eq:V-X=l}), where the $m_Q'$ is canceled in the potential. It turns to be the exact solvable case.\ \\

\textbf{Example 2. $A=\lambda$, $B=0$, $C_1=0$, $C_2=-i$}

In this case, we have $X(x)=\sqrt{\lambda}x$. For the transformation, we get the QES potential
\begin{eqnarray}\label{eq:V-X=x}
\mathcal{V}(r)&=&\frac{2m\omega^2}{\lambda}\left(\textrm{sech}(\Upsilon(r))-\textrm{tanh}(\Upsilon(r))\right)^2+\frac{\hbar^2}{8mr^2}\left[1+2\lambda r^2-4{m_Q'}^2(1+\lambda r^2)\right]\\
&&+\frac{\lambda \hbar^2}{2m}\left\{{m_Q'}\left(5{m_Q'}-3\delta\right)\textrm{sech}^2(\Upsilon(r))+\left[-2+2{m_Q'}(5+4{m_Q'})-5\delta\right]\textrm{sech}(\Upsilon(r))\textrm{tanh}(\Upsilon(r))\right.\nonumber\\
&&\ \ \ \ \ \ \ \ \ \ +\left.\left[6+5{m_Q'}(2+{m_Q'})+5(1+{m_Q'})\delta\right]\textrm{tanh}^2(\Upsilon(r))\right\}\nonumber
\end{eqnarray}
and the ground-state wave function
\begin{equation}\label{eq:f-X=x}
\psi_0(r)=\mathcal{N}_0(\lambda r^2)^{-1/4}(1+\lambda r^2)^{-1/2}\left[\textrm{sech}(\Upsilon(r))\right]^{\beta/\lambda}\exp(-\frac{\gamma}{\lambda}\textrm{gd}(\Upsilon(r)))
\end{equation}
where $\textrm{gd}(x)=2\arctan e^x-\frac{\pi}{2}$ is the Gudermannian function.

The excited state wave functions can be gotten by the raising operator $b^\dagger$ in Ref. \cite{wang} $\phi_n(x)=(b^\dagger)^n\phi_0(x)$ and the transformation as $\psi_n(r)=g(r)\phi_n(x(r))$.

For the excited state, however, the two-dimensional generalized Higgs model has exactly solutions only in the condition of angular parameter $m'$ equaling to $m_Q'$. In other angular parameter cases, it cannot be solved. Therefore, it is a quasi-exactly solvable model.

%In this case, we have
%\begin{equation}\label{eq:V-X=x2}
%\mathcal{V}(r)=\frac{\hbar^2\left[\beta+2\beta\gamma+3\gamma\lambda+\gamma(1+\gamma)\csch^2(\Upsilon(r))+\beta(\beta+3\lambda)\sinh^2(\Upsilon(r))\right]}{8m\left[1+\lambda\sinh^2(\Upsilon(r))\right]}
%\end{equation}
%and the energy is
%\begin{equation}\label{eq:E-X=x2}
%E_n=-\frac{\hbar^2}{2m}3\lambda(n-\frac{\beta}{3\lambda})^2+\frac{\hbar^2}{2m}\frac{\beta(\beta+3\lambda)}{3\lambda};
%\end{equation}

%\textbf{Example 3. $X=\cos\left(\Theta(x)\right)$}\ \\

%\textbf{Example 4. $X=\cos\left(3\Theta(x)\right)$}\ \\

\section{Conclusion}

In this paper, we get the transformation between the generalized CRS model and the radial Higgs model. Based on this transformation, we construct a series of the quasi-exactly solvable generalized Higgs models. Applying this method, we can construct a series of two-dimensional quasi-exactly solvable model by connecting a dimensional exactly solvable model with the radial part of a two-dimensional model. Thus, we can get a plenty of two-dimensional quasi-exactly solvable models finally. Furthermore, based on the result presented in this paper, we can also discuss
the eigenproblem of a quantum system, e.g., the counting function of the
eigenvalue which is discussed in Ref. \cite{dai2009number}%
\cite{dai2010approach}. Moreover, we can also discuss the relation between the
operator algebra and the corresponding statistical distribution
\cite{dai2012calculating}. Our result can also be applied into other
statistical mechanics problems \cite{song2014}.

\section*{Acknowledgements}

We thank Da-Bao Yang, Fu-Lin Zhang, Tong Liu, Wen-Du Li, Yu-Zhu Chen and Chi-Chun Zhou for their helpful discussion.This work is supported by the National Natural Science Foundation of China (Grant Nos. 11175089 and 11075077), the National Basic Research Program of China (Grant No. 2012CB921900)

%\bibliography{ConFid_flzhang_rsb}
%\bibliographystyle{apsrmp4-1}
\bibliographystyle{JHEP} %参考文献的风格
\bibliography{QESH} %参考文献文件
\nocite{*} %没有被引用的文献也被列出

\end{document}